# ECHO-3DHPC: Advance the performance of astrophysics simulations with code modernization

*Matteo Bugli*, PhD, Astrophysicist, CEA Saclay, Département d'Astrophysique (DAp)/DEDIP, Gif-sur-Yvette, France

*Luigi Iapichino,* PhD, Scientific Computing Expert, Leibniz Supercomputing Centre of the Bavarian Academy of Sciences and Humanities, Garching b. München, Germany

*Fabio Baruffa*, PhD, Technical Consulting Engineer, Intel Corporation, Feldkirchen, Germany

Accurate and fast numerical modeling is essential to investigate astrophysical compact objects (such as neutron stars and black holes) which populate every galaxy in our universe and play a key role in understanding astrophysical sources of high-energy radiation. It is particularly important to study how hot plasma accretes onto black holes, as this physical phenomenon can release huge amounts of energy and power up some of the most energetic objects in the cosmos (such as gamma-ray bursts, active galactic nuclei and x-ray binaries). The numerical studies that aim at investigate the accretion scenario end up being rather computationally expensive, especially when the understanding of the fundamental physical mechanisms requires the exploration of multiple parameters. For this reason the modeling of relativistic plasmas greatly benefits from an efficient exploitation of HPC resources.

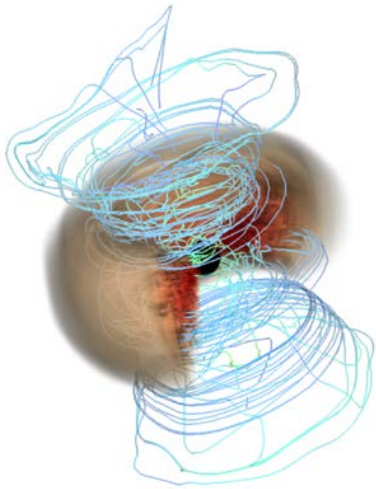

Figure 1: Volume rendering of the mass density of a thick accretion torus (dense regions in red, rarefied envelope in brown, velocity flow as streamlines), simulated with the ECHO-3DHPC code [3].

ECHO-3DHPC is a Fortran application developed to solve the equations of magneto-hydrodynamic in General Relativity, using 3D finite-differences method and high-order reconstruction algorithms [5]. It is originally based on the code ECHO [1], and in its most recent version it employs a multidimensional MPI domain decomposition scheme that shows very high scalability with large number of cores, as demonstrated during the 2017 Scaling Workshop at LRZ, where 65 536 Xeon cores of SuperMUC Phase 1 have been utilized [2].

Recent code developments introduced multi-threading parallelism using OpenMP*, which allows to scale the application beyond the current limit set by MPI communication overhead. With the help of the Intel® Software Development Tools, like Fortran compiler and Profile-Guided Optimization (PGO), Intel MPI library, VTune™ Amplifier and Inspector we have investigated the performance issues and improved the application scalability and the time to solution.

**Using Intel Compiler for Performance Optimization**

A first and mostly overlooked ingredient for performance optimization is a better use of compiler features, like the Profile-Guided Optimization (PGO). PGO facilitates the developer's work in re-ordering code layout to reduce instruction-cache problems, shrinking code size, and reducing branch mispredictions. The procedure of its use consists of three steps:

- compile your application with **-prof-gen** option, which instruments the binary;
- run the application generated in the previous step, to generate a dynamic information file;
- recompile your code with the option **-prof-use**, to merge the collected information and generate an optimized binary file.

Performance measurements show an improvement up to 15% on the time to solution for the largest run with 16 384 cores, see Table 1.

| Grid size | # of cores | Intel compiler 18 [sec/iter] | Intel compiler 18 with PGO [sec/iter] |
| --- | --- | --- | --- |
| $512^3$ | 8 192 | 1.27 | 1.13 |
| $512^3$ | 16 384 | 0.66 | 0.57 |
| $1024^3$ | 16 384 | 4.91 | 4.22 |

Table 1: Hardware configuration: dual-socket Intel Xeon E5-2680 v1 (code-named Sandy-Bridge) @ 2.7GHz, 16 cores per node. Software: SLES11; Intel MPI library 2018.

**OpenMP optimization using VTune and Inspector**

Recently a multi-threaded parallelization using OpenMP has been implemented in ECHO-3DHPC. This allows to reduce the number of MPI tasks by using OpenMP threads within a CPU node, thus relieving the pressure exerted by MPI communication in multi-node runs.

Performance optimization is based on single-node tests on a dual-socket Intel Xeon E5-2697 v3 (code-named Haswell) with a total of 28 cores @ 2.6 GHz. The command line for generating the *HPC performance characterization* analysis Intel® VTune™Amplifier is:

**amplxe-cl -collect hpc-performance -- <executable and arguments>**

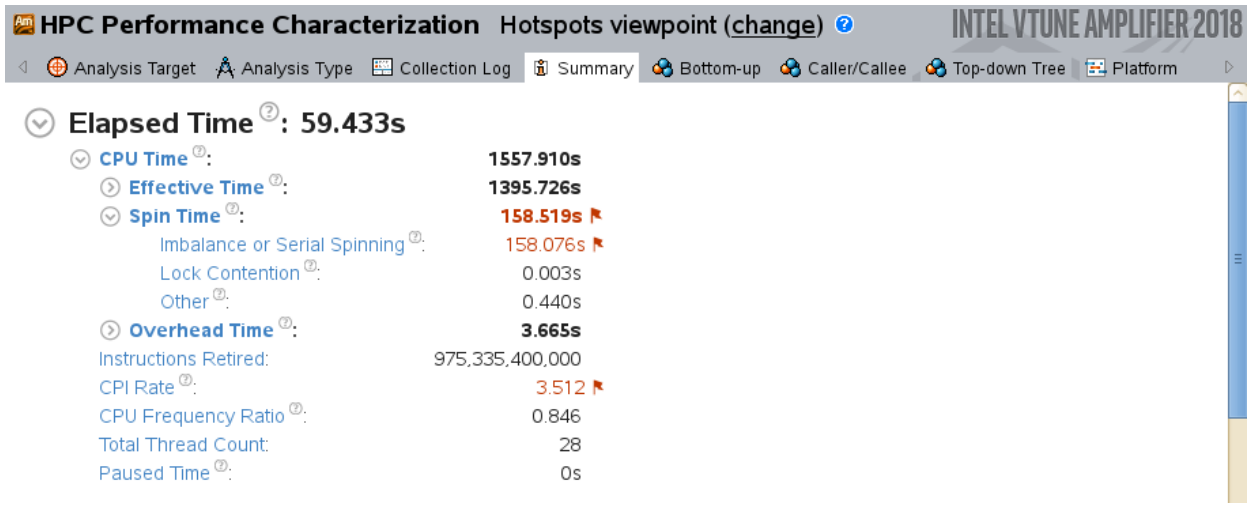

Figure 2: Screenshot from the VTune *HPC Performance Optimization, Hotspots viewpoint* analysis for the baseline version of the ECHO-3DHPC code.

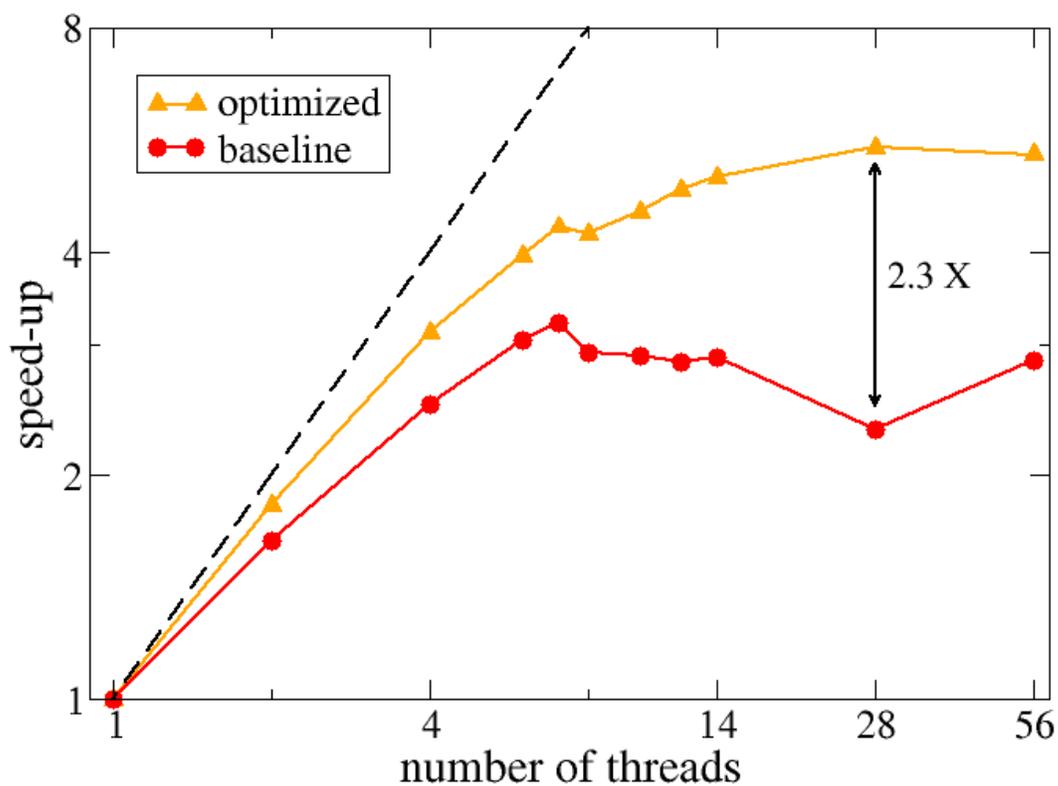

Figure 3: Parallel speed-up at node level as a function of the number of threads (the last point refers to hyperthreading). The red and orange lines refer to the baseline and optimized code versions. Ideal scaling is indicated by the dashed black line.

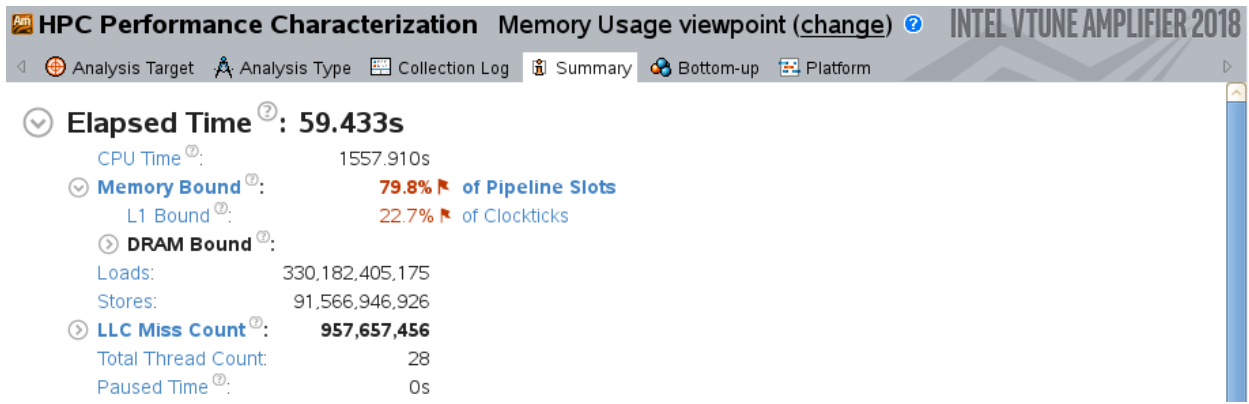

Figure 4: Screenshot from the VTune *HPC Performance Optimization, Memory usage viewpoint* analysis for the baseline version of the ECHO-3DHPC code.

The analysis of the performance of the baseline code version shows the following bottlenecks:

- a large fraction of the CPU time is classified as "Imbalance or serial spinning" (Figure 2). This is caused by insufficient concurrency of working threads, due to a significant fraction of sequential execution;
- as a consequence of the previous point, the node-level scalability suffers with a large number of threads (Figure 3, red line);
- the Memory access analysis of VTune Amplifier shows that the code is memory bound, with about 80% of execution pipeline slots stalled because of demand of memory loads and stores (Figure 4).

Besides the performance considerations above, the correctness of the OpenMP implementation (only recently introduced in the code) is worth being investigated with Intel® Inspector. Intel® Inspector is an easy-to-use memory and threading error debugger for C, C++, and Fortran applications. It does not require special compilers or builds and uses a normal debug or production build. This kind of analysis type might have impact on the application runtime due to instrumentations overhead.
The command line to investigate and correct threading errors is:

**inspxe-cl -collect=ti3 -- <executable and arguments>**

and indeed the analysis shows a number of data races (Figure 5). These are caused by multiple threads accessing the same shared memory location, which might create also a non-deterministic behavior.

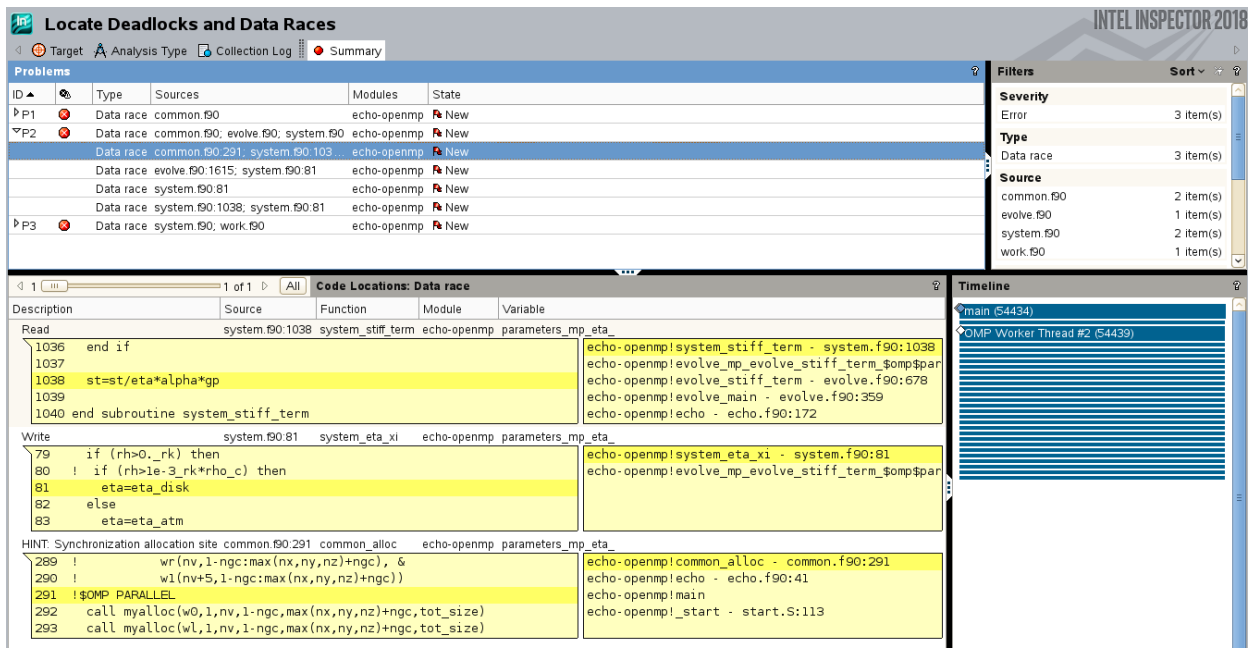

Figure 5: Screenshot of the Inspector *Locate Deadlocks and Data Races* analysis for the baseline version of the ECHO-3DHPC code.

Based on the above findings, we have increased the number of parallel regions, changed the workload scheduling to *guided* and applied PRIVATE clauses to selected variables in OpenMP *PARALLEL DO* loops, thus removing the data races previously detected by Inspector. The effect of these optimizations can be summarized as follows:

- on 28 threads the optimized code has a performance improvement of 2.3x (Figure 3) with respect to the baseline version, and a parallel speed-up of 5.5x. This is in agreement with the prediction of Amdahl's law for a parallel application with a fraction of 15% of sequential execution, as currently measured in the code using Intel® Advisor tool (more details in[4]);
- the spin time decreases by 1.7x;
- the code is still memory bound, but the fraction of stalled pipeline slots has slightly decreased to 68%.

**Overcoming the MPI communication bottleneck**

The effect of the OpenMP optimization described above can be appreciated also in the analysis of hybrid (MPI-OpenMP), large-scale runs. In Figure 6 the scalability of the pure-MPI and of the hybrid code are compared, in terms of time per iteration (lower is better).

We report here the results for the best hybrid configuration, with 4 MPI tasks per compute node and 7 OpenMP threads per task. At small number of compute nodes, the MPI-only version of the code (blue line) has better performance. The picture changes at larger numbers of nodes, where the scalability starts degrading because of the MPI communication overhead. The problem size per MPI task is not large enough to compensate the communication time (*communication is dominating the computation*). The number of nodes with the best performance (the lowest point in the scaling curve) is moved to the right of Figure 6 from 32 to 128 nodes, corresponding to 896 and 3 584 cores, respectively, for the hybrid MPI-OpenMP code (green line), thus enabling a more effective use of larger HPC systems. Comparing the runs with best performance, the time spent in MPI communication has decreased by 2x in the hybrid configuration. Additional optimization of the code will allow the use of even more OpenMP threads per node, further relieving the communication overhead.

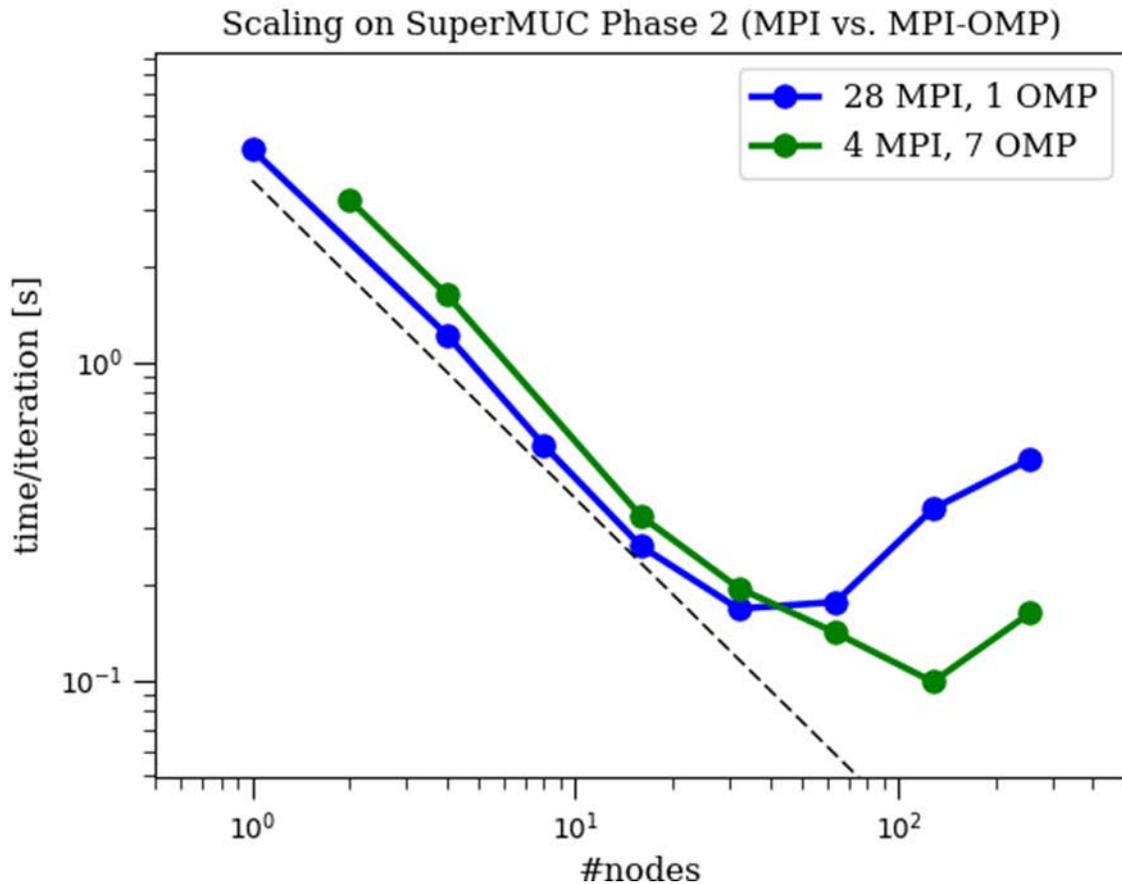

Figure 6: Scalability plot for the pure-MPI (blue line) and the hybrid, MPI-OpenMP (green line) code versions. The dashed lines represent the ideal strong scaling. We remind here that the Haswell nodes used for these runs have 28 cores.

**Final remarks**

We presented recent developments in the parallelization scheme of ECHO-3DHPC, an efficient astrophysical code used in the modelling of relativistic plasmas. The code's new version employs an optimized hybrid MPI-OMP algorithm, exhibiting good scaling properties on more than 65 000 cores and enabling a more efficient exploitation of HPC resources and hence faster time to solution.